\documentstyle[aps,twocolumn,epsf]{revtex}

\newcommand{\beq}{\begin{equation}}
\newcommand{\eeq}{\end{equation}}
\newcommand{\bea}{\begin{eqnarray}}
\newcommand{\eea}{\end{eqnarray}}
\newcommand{\bean}{\begin{eqnarray*}}
\newcommand{\eean}{\end{eqnarray*}}
\newcommand{\lab}[1]{\label{#1}}
\newlength{\smallwidth}
\newlength{\red}
\newcommand{\s}{\sigma}

\newcommand{\ta}{\tau}

\newcommand{\ph}{\varphi}

\newcommand{\la}{\lambda}

\newcommand{\Om}{\Omega}
\newcommand{\Si}{\Sigma}
\renewcommand{\a}{\alpha}
\newcommand{\g}{\gamma}
\renewcommand{\b}{\beta}

\renewcommand{\d}{\delta}

\newcommand{\cT}{{\cal T}}
\newcommand{\cA}{{\cal A}}

\newcommand{\cE}{{\cal E}}
\newcommand{\cR}{{\cal R}}
\newcommand{\cO}{{\cal O}}
\newcommand{\cF}{{\cal F}}

\newcommand{\cX}{{\cal X}}

\newcommand{\cV}{{\cal V}}
\newcommand{\cC}{{\cal C}}
\newcommand{\cS}{{\cal S}}
\newcommand{\bra}{\left<}
\newcommand{\ket}{\right>}

\newcommand{\hL}{\hat L}

\newcommand{\di}{\Diamond}

\newcommand{\nn}{\nonumber}
\newcommand{\1}{\mbox{1 \hspace*{ -1em} 1  }}

\newcommand{\9}{\partial}
\newcommand{\de}{\delta}

\newcommand{\tr}{\mbox{tr}}

\newlength{\oldarraycolsep}

\newlength{\abswidth}
\begin{document}

\title{
 \vskip-1.0cm{\baselineskip14pt}
  \centerline{\normalsize\hskip12.5cm HD--THEP 98--40}
  \vskip0.5cm
Time evolution of observables out of 
       thermal equilibrium. }

\author{Herbert Nachbagauer}

\address{Institut f\"ur Theoretische Physik, Universit\"at Heidelberg, \\
Philosophenweg 16, D-69120 Heidelberg, Germany}

\date{Talk given at the 5th International
Workshop on Thermal Field Theories\\ and Their Applications, Regensburg,
10--14 August 1998}

\maketitle

\begin{abstract}

We propose a new approximation--technique to deal with the exact
macroscopic integro--differential evolution equations of statistical
systems which self--consistently accounts for dissipative effects. 
Concentrating on one and two point equal--time correlators, we develop
the self--consistent method and apply it 
to a scalar field theory with
quartic self--interaction. We derive the effective equations of motion
and the corresponding macroscopic effective
Hamiltonian. Non--locality in time appears in a natural way 
necessary to account for entropy generating processes. 

\end{abstract}

\pacs{05.70.Ln, 11.10.Wx, 11.15.Tk }

\narrowtext


\section{Introduction}

The question of how and if a statistical infinite dimensional quantum
system
thermalises is one of the fundamental problems of statistical
mechanics, and  has recently attracted much attention in connection
with problems in cosmology and relativistic heavy ion physics. 
We discussed some related issues in a recent work 
\cite{nach98/1} for a zero dimensional quantum mechanical model. 
There, we concluded that a safe and 
consistent method to investigate on
time evolution of macroscopic observables is to apply the projection
operator technique. In this work, we systematically apply the new 
self--consistent approximation technique anticipated in \cite{nach98/1} 
to a scalar field theory with quartic self--interaction. 
For systematic reasons, let us therefore briefly review the line of 
arguments leading to that concept. 
 
In statistical theory, a level of observation is 
defined by a set of Hermitian so--called
relevant operators the expectation values of which are the macroscopic 
observables of the ensemble under consideration. 
The dynamics is defined by the microscopic
Hamiltonian which enters in the exact integro--differential equations
of motion 
for the macroscopic observables, as derived by the
projection--operator technique \cite{proj}. These equations solve 
the complete BBGKY-hierarchy in integral form and are closed in 
the observables under consideration. Correlators not 
expressible in terms of the basic observables are 
eliminated by the projection procedure but leave their imprints 
implicitly in the defining equation of a projected
evolution operator. The problem of resolving the infinite 
set of differential equations of the BBGKY--hierarchy, and subsequent
reduction to the macroscopically relevant quantities, has thus been
converted into approximating that operator in a self--consistent 
way. Only in some particular cases, as for effectively non--interacting 
systems defined by a Hamiltonian which contain polynomials of 
at most second order in position and momentum operators,
the projected time evolution operator can be integrated 
without explicit dependence on projectors. 
A side--condition for exact
integrability is that also the level of observation must be
chosen to be at most of second order (quadratic level of
observation). 
The physical reason for exact integrability is 
dynamical closure, i.e. the fact that time evolution driven by effectively
non--interacting Hamiltonians does not lead out of the linear space
spanned by the operators of a quadratic level of observation.
Typical examples of this type are time dependent mass--parameters or
QED without quantized fermions, but including classical external sources. 

For the interacting case, 
the expansion of the projected evolution operator into a coupling parameter
labeling the interaction part is discussed extensively in the
literature \cite{proj}. However, that expansion scheme turns out
to be inconsistent with the projected dynamics at second order even 
when we introduce an effective quadratic 
 micro-dynamical Hamiltonian and expand the 
projected time evolution operator around it.  That was demonstrated
explicitly in a zero-dimensional toy model recently \cite{nach98/1}. 
The  break--down of that kind of expansions becomes apparent if one 
integrates the system for times large compared to the inverse coupling
of the interaction part. The physical reason for that may best be
illustrated  by considering the 
uncertainty product expressible in terms of the quadratic observables. 
Even for effectively non--interacting Hamiltonians with arbitrary time
dependent coefficients, the uncertainty -- which essentially is the volume of
phase space cells -- is an exact constant of motion. However,
the dissipative part of the equations of motion for the observables
does not respect constant uncertainty, and evolves it to numerically
large values at secular time scales. On the other hand, 
increasing uncertainty and increasing phase space volume of the corresponding
macroscopic dynamics is an essential feature of non--equilibrium 
statistical systems. Moreover, the uncertainty product is directly 
related to the relevant entropy of a quadratic level of observation. 
We concluded that any effective microscopic Hamiltonian is intrinsically
incompatible with time variant entropy and cannot be
used to approximate the projected evolution operator. One necessarily
has to generalize to an effective Liouvillian acting in the
space of relevant operators and expand around its associated
so--called reduced time evolution operator. 

The task of the present investigation is to derive the effective 
equations of motion by approximating the projected
evolution operator by its reduced counterpart to the first non-trivial
order.  We also construct the
corresponding effective Hamiltonian which generates the macroscopic 
equations of motion.

\section{Definition of the problem}

(i) A mixed state of a quantum system (configuration), both in zero
dimensional quantum mechanics as well as in quantum field theory, 
is described by a density operator, which, in order to allow a 
probability interpretation, must be a hermitian trace-class operator with 
positive eigenvalues. It is an intrinsic feature of quantum 
systems that the density matrix is fictive in the sense that only 
its diagonal elements correspond to physical probabilities, the 
other dependencies being phases which enter in expectation
values via interference effects\footnote{That is the fundamental 
difference to classical phase space averages.}.  Alternatively, 
one may characterize a configuration 
by the expectation values of hermitian operators. A generic 
set of those representing a complete set of observables is 
given by the mutually orthogonal hermitian projectors constructed 
from the eigenvectors of the density matrix. A complete set of 
(not necessarily commuting) observables contains all information 
about the density matrix such that any observable can be expressed 
in terms of the complete set. 

(ii) The system may be assigned a dynamical structure. Motion is 
defined as a sequence of possible states having a constant 
expectation value of the Hamiltonian operator, the energy of the 
configuration. Quantum mechanical time parameterizes those 
configurations which are assumed to have time--independent 
probabilities and phases for autonomous systems. The time evolution 
generated by the micro-dynamical Hamiltonian can be expressed by 
first order differential 
operator--equations in time for the density matrix  (von~Neumann 
equation) in the Schr\"odinger representation,
\beq
i \frac{d}{dt} \rho(t)  = [H, \rho(t) ], 
\eeq
together with an initial condition $\rho(0)$.
Hermitian Hamilton--operators generate unitary time evolution which in
turn is necessary to allow for a  probability interpretation. 
Once the problem of time evolution of the 
mixed state is solved for a given initial configuration, observables can be 
calculated as expectation values from the evolved density matrix. 

(iii) The statistical description of a system is based on a reduction 
procedure selecting an in general much smaller number of (macroscopic) 
observed quantities out of the complete set of observables. 
This subset defines the level of observation (description),
characterized by the set of relevant operators
$\cE = \lbrace \cF_\nu \rbrace  . $ 
The process of reduction from the density operator to 
the set of the so--called relevant quantities in general involves loss of 
information. Here, we will concentrate on levels of observation that
once chosen, will be kept fixed during time evolution. That constraint
can be relaxed too if necessary \cite{balian86}.

(iv) The reduction can be dynamically trivial if it commutes with time 
evolution. In that case of dynamical closure, the Liouvillian maps the
operators of the 
level of description on a linear combination of them. The observables of 
a level of description at a certain time are sufficient to determine them at 
any time and evolution induces neither information gain nor information 
loss at the level of observation, the associated entropy being constant.
On account of the canonical commutation relation, the most general 
Hamiltonian admitting a finite dimensional dynamically closed level of 
observation can contain constant, linear and quadratic expressions in 
position and momentum operators\footnote{Regardless of additional 
dependencies on other c-number quantities, including time and classical 
field strength, we call it free Hamiltonian.}. The corresponding 
dynamically closed sets of observables correspond to sums of polynomials of 
finite order in the canonical operators.  
If the reduction is non-trivial, one can extend the level of observation
to render it trivial. That may involve an infinite number of operators in 
which case the system is a truly interacting one, and the 
only dynamically closed set of operators corresponds to a complete 
set of observables.  

(v) For truly interacting dynamical systems, the complete initial 
density matrix influences observables at later times. Its definition
calls for an additional principle to construct it from the reduced set of 
initial data. Information theory proposes to apply Shannon's   
theory of entropy \cite{shan49} to that physical problem. Jaynes' principle 
of maximum entropy \cite{jayn89} fixes the generalized canonical 
density operator as initial condition. It can be shown not to contain more 
information than the initial set of observables. We want to point out that 
this choice is the statistically most probable, but the underlying concept of 
ensemble averages of identical systems evolving from variant
initial preparations does not imply actual the preparation of the 
physical system in that state.  

The reduced description of a quantum system is achieved by passing
from the density matrix to a statistical operator characteristic for
the ensemble under consideration. The best guess for that operator
compatible with the principle of maximum entropy has the functional form 
$$ \cR=\exp( {-\mu_\nu \cF_\nu}) . $$
It contains exactly the same amount of information on the system as
the statistical observables $g_\nu = \tr \left( \rho  \cF_\nu \right)
= \tr \left( \cR \cF_\nu \right) $, and  maximizes the relevant
entropy functional of the level of observation,
\beq  S_{\cE} = - \tr \left( \cR \log\cR \right) = \mu_\nu g_\nu . 
\label{srel} \eeq
Reducing the description of a quantum mixed state  to the evolution of
the accompanying statistical operator $\cR$ requires to derive a
modified evolution equation which does not lead out of the level of
observation. The problem basically boils down to introducing suitable 
projection operators and the study of their associated time evolution.

\begin{figure}[t]
\epsfxsize=11cm
\vskip-1cm
\centerline{\epsfbox{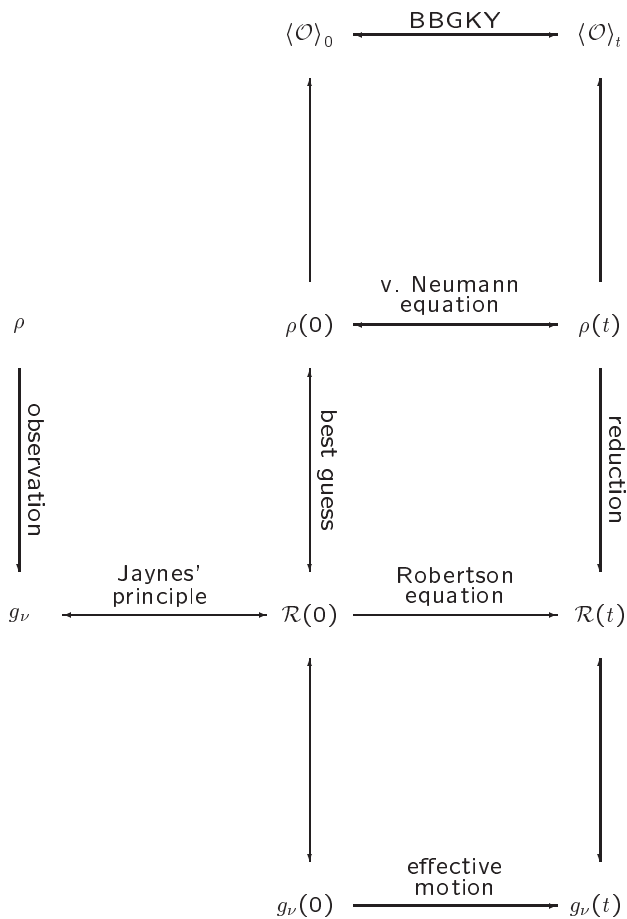}}
\caption{Overview of relations on time evolution of mixed states
and statistical time evolution. The arrows indicate the sense of
deduction.} 
\label{fig1}
\end{figure}

An overview of the systematics is given in Fig.\ (1). Here, we adopt
the Robertson equation approach. Alternatively, one may solve the
BBGKY--hierarchy which, however, is problematic since it involves an
ad hoc truncation procedure. Problems with that approach 
have been discussed extensively elsewhere \cite{nach98/1}. 

\section{Projection Operator Technique}

We briefly review some basic terms of the projection operator approach
\cite{proj,balian86}
serving as starting point of the present investigation. There, one
defines the level of observation by the finite set of
Hermitian operators $\cE=\lbrace \cF_\nu \rbrace $ including
$\cF_0=\1$ for convenience. The accompanying canonical density 
operator defined by
$\cR(t) = \exp ( - \mu_\nu (t) \cF_\nu )$ contains time dependent
Lagrange multipliers $\mu_\nu(t) $ which are functions 
of $g_\nu(t)$ such that $g_\nu(t) = \tr \left( \cF_\nu \rho(t) \right)=
\tr \left( \cF_\nu \cR(t) \right) $. 
We emphasize that other than the density matrix $\rho(t)$,
$\cR(t)$ in general does not evolve according to a Schr\"odinger
equation.

For the corresponding operator expectation values $g_\nu(t) $, 
the closed exact equation (Robertson equation) of motion is found 
to read  
\bea
&& \frac{d}{dt}  g_\nu(t)
= -i \, \tr \left( \cF_\nu  L \circ \cR(t) \right)  \nn \\ 
 && \quad - \int_{0}^t \! dt' \, \tr \left( \cF_\nu L \circ T(t,t') \circ Q(t')
\circ L \circ \cR(t') \right) . \label{eom1}
\eea 
Here $L$ stands for the Liouvillian induced by the micro-dynamical
Hamiltonian, and the action of $L$ on some operator $\cX$ is defined
by\footnote{The expression  $ A \circ B$ in general 
denotes the action of operator $A$ on operator $B$. There may be
no representation of $A \circ {}$ in terms of a commutator.}
$L\circ \cX = [H,\cX]$.

The r.h.s.\ of (\ref{eom1}) introduces two new symbols which call for
further explanation. The projector $Q(t)$ is defined by 
\beq
\tr ( \cO_1 Q(t) \circ \cO_2 ) = \tr  ( \cO_1  \cO_2 )  -  
 \frac{\9 \tr ( \cO_1 \cR(t) )  }{\9 g_\nu(t) }  \,  \tr ( \cF_\nu
\cO_2 ) ,
\label{proj}
\eeq
for any pair of operators $ \cO_1 , \cO_2$. In the particular case of
$\cO_1$ being in the linear space of the level of observation $\cE$, 
the rhs of Eq.\ (\ref{proj}) vanishes.  

The non-unitary projected evolution operator $T(t,t')$ is a solution of 
\bea
\frac{\9}{\9 t} T(t,t') &=& - i Q(t) \circ L \circ T(t,t') \qquad
\mbox{and}  \nn \\   
\frac{\9}{\9 t'} T(t,t') &=& i  T(t,t') \circ  Q(t') \circ L  \nn 
\eea
with initial condition $T(t,t) = 1 $. It will turn out to be the 
key problem to construct this operator from its definition for the
particular Hamiltonian under consideration. Then, all trace expressions 
at the r.h.s.\ of (\ref{eom1}) can at least in principle be expressed
in terms of the $g_\nu$, and the system is a closed integro-differential
equation in the c-number expectation values.  

In this investigation, we will complete $\cE$ such that 
the action of the Hamiltonian induced Liouvillian can  be split 
into $L=L_0+L_I$, and the free part be dynamically closed 
with respect to the level of observation, 
$L_0 \circ \cF_\nu  = [ H_0 , \cF_\nu ] = \Om_{\nu\mu} \cF_\mu$.
In that case, in the integral part of Eq.\ (\ref{eom1}), 
the complete Liouvillian can be replaced by
$L_I$.  

Expectation values of operators 
which cannot be represented as linear combinations of the level of
observation get additional contributions to their 
$\cR$-averages,
\bea
 \tr (\cO \rho(t) ) &=& \tr \left( \cO \cR(t) \right) \nn \\
&-&  i \int_0^t dt' \tr \left( \cO T(t,t') \circ Q(t') \circ L \circ
\cR(t') \right) , \label{r-aver}
\eea
where again the dynamically closed part in $L$ does not contribute to the 
integral. 

\section{Projected evolution} 
The key problem is to find a consistent approximation scheme to solve
the defining equation for the projected evolution operator $T(t,t')$. 
That scheme --- order by order -- 
must (i) be compatible with the Robertson equation and (ii)
conserve exact constants of motion. In particular, the expectation
value of the Hamiltonian operator being a generic constant of motion 
must be conserved which is a non--trivial condition since, due to
relation (\ref{r-aver}), that generally involves integral terms non--local in
time. The query for a consistent approximation scheme naturally raises the
question whether there exists an effective micro-canonical Hamiltonian
which may serve as a zeroth--order approximation being a staring point
of an expansion. 

A fist approach to that problem is to study the small $\la$  expansion of
$H = H_0 + \la H_I$. If one evaluates the corresponding expressions to
second order in the coupling and solves the resulting system of
integro--differential equations of motion, one finds that the
expansion breaks down at typical secular time scales of the order 
$t \sim 1/\la $ \cite{nach98/1}. 

An improvement of that expansion scheme in equilibrium theory can be
achieved by resummation. In our case, that amounts to introduce an
effective Hamiltonian. In particular, one may try to 
make an quadratic Ansatz of the
form $$H_{\mbox{eff}} = a(t;x,y) \Pi(x) \Pi(y) + b(t;x,y) \Phi(x) \Phi(y)  +
\ldots  $$
and determine the time dependent coefficients by
self--consistency requirements. However, it turns out that even that
improved scheme fails to meet the consistency conditions (i,ii)
at secular time scales \cite{nach98/1}. The physical reason for that
is the presence of memory effects. In particular, although the effective
Hamiltonian correctly describes time evolution for short times $t \ll
1/\la$ in an
appropriate way, the time evolved quantity $H_{\mbox{eff}}$ does not
coincide with  effective Hamiltonian constructed by the consistency
requirements at later times. We
conclude that only the enveloping dynamics is an effective
Hamiltonian one. Furthermore, effective Hamiltonian dynamics
automatically gives rise to an additional constant of motion related
to the uncertainty product of the configuration. On account of the
uncertainty--entropy relation for quadratic observables, that results
in conservation of the relevant entropy contradicting constant
information loss inherent in the dissipative time 
evolution of statistical
systems. Also, locality in time conflicts with the presence of memory
terms, which get important at large time scales. Third, an effective
Hamiltonian automatically induces a unitary effective time evolution 
operator which again conserves observable information. 
The conclusion is that there generally exists no effective
micro-dynamical Hamiltonian dynamics which is compatible with the
Robertson--equation and conservation of energy. We thus have to
completely abandon the query for  micro-dynamical 
Hamiltonians in favor of a the more general concept of 
Liouvillian dynamics.  

The key idea is to formally introduce an effective
evolution operator which evolves the accompanying statistical
operator $\cR(t)$ in a form--invariant way and expand the projected
evolution operator around that reduced time evolution. 

The reduced time evolution super--operator $\cV(t,t')$ is uniquely
defined by two demands. Firstly, it does not lead out of the 
linear space of 
the relevant operators (observables). Secondly, it evolves the
statistical operator,
\beq 
\cR(t) = \cV(t,t') \circ \cR(t').
\eeq
That condition already is very restrictive. Due to the particular form of the
accompanying statistical operator, the differentiated equation
suggests to write a solution as a time-ordered exponential
\beq
\cV(t,t')  = \cT \exp \left( -i \int_{t'}^{t} d\ta \hL(\ta) \circ 
{} \right) .\lab{vv}
\eeq
Since the derivative $\9_t$ obeys the Leibnitz product--rule,
the effective Liouvillian--Operator $\hL(t)$ has to have the algebraic
property of a derivation,
$ \hL(t) \circ ( A B  ) =  ( \hL(t)\circ A ) B + A ( \hL(t)\circ B )$. 
We want to point out that in general this Liouvillian is not an
anti--hermitian operator, $ \hL \neq - \hL^\dag $ and thus cannot be
cast as a commutator with some effective Hamiltonian, 
$\hL \neq  [ \hat H,.\,] . $

Furthermore, on account of linearity in the level of observation, 
the reduced time evolution can be expressed in terms of a matrix
evolution operator, $ \cV(t,t') \circ \cF_\a = \cV_{\a\b}(t,t') 
\cF_\b$. The reduced time evolution operator still possesses the
property of transitivity,
$ \cV(t,t') \circ  \cV(t',t'') = \cV(t,t'') $. 

In practical calculations, it turns out to be convenient to act with 
the reduced time evolution operator to its left which amounts to
define the adjoint action induced by the trace as inner product by
\bea  \tr \left( \cO \cV(t,t') \circ \cR(t') \right) &=& 
\tr \left( (\cV^\dag (t,t') \circ \cO ) \cR(t') \right) \nn \\ 
 &=& 
\tr \left( \cO(t,t')  \cR(t') \right) .
\eea
The adjoint evolved operator $\cO(t,t')$ is the generalization of the
free 
Heisenberg operator to $\cO$ for non--Hamiltonian time evolution. 
Due to the particular properties of $\cV(t,t')$, one can also express
time evolution of the generalized Heisenberg
operators of the basis by means of a matrix multiplication,
$ \cF_\a(t,t') = \cV^\dag (t,t') \circ \cF_\a = 
\cV^\dag_{\a\b} (t,t') \cF_\b .$
The matrix elements are restricted by the conditions 
\beq
g_\a(t) = \cV^\dag_{\a\b} (t,t') g_\b(t')  . \lab{s1} 
\eeq
In addition to that, the property 
\beq
\cV^\dag (t,t') \circ ( \cF_\a \cF_\b ) = 
\left( \cV^\dag (t,t') \circ \cF_\a \right)  \left( \cV^\dag (t,t')
\circ \cF_\b \right) \label{s2}
\eeq 
which is a direct consequence of $\hL(t)$ being a derivation, further 
restricts the number of independent elements in the evolution matrix
if some symmetrised products $\lbrace \cF_\a , \cF_\b \rbrace $ are
also in the operator basis of the level of observation.
In general, however, those conditions are not sufficient to fix the
evolution matrix completely. 

The self--consistent method now requires to expand the projected evolution
operator in terms of the reduced time evolution operator.
It is part of our strategy to exploit the remaining freedom in the 
evolution matrix to satisfy consistency with the Robertson evolution
equations, and possible conserved quantities. As will be shown in the
case of the local and bilocal equal--time correlators as observables, 
that fixes the dissipative evolution matrix in a consistent way. 

\section{Reduced evolution operator for one and two point correlators} 

In order to define the ensemble, a particular set of observables must
be chosen. Among various choices of the level of observation, the special
case of quadratic observables plays a particular r\^ole. This set of 
Hermitian operators that can be constructed out of the  field operators and its
canonically conjugate momentum, $\lbrace 
\1$, $\Phi(x)$, $\Pi(x)$,  $\Phi(x) \Phi(y)$,
$ W(x,y) = \Phi(x) \Pi(y) + \Pi(y) \Phi(x) $, $ \Pi(x) \Pi(y)
\rbrace ,$
and give rise to a quasi Gaussian bilocal accompanying statistical
operator. This particular choice of the level of observation renders
possible to apply a generalized Wick theorem in
the evaluation of weightened quantities $\bra \cO \ket_t = \tr
\left( \cO \cR(t)\right) $. 
Then expectation values of polynomials in $\Phi,\Pi$ can 
always be expressed as products of contracted pairs of the macroscopic
observables 
$\ph(x|t) = \bra \Phi(x) \ket_t , \, \pi(x|t) = \bra \Pi(x) \ket_t
$ , and the correlators
\bea
\ph_2(x,y|t) = \bra \Phi(x) \Phi(y) \ket_t , & &  \pi_2(x,y|t) = 
\bra \Pi(x)\Pi(y) \ket_t , \nn  \\   w(x,y|t) &=& \bra W(x,y) \ket_t .
\eea

We now construct the reduced time evolution operator for this level
of observation. Since $\1,\Phi,\Pi$ already represent a basis out of which
the quadratic quantities are formed, it suffices to investigate on the
evolution matrix elements contained in  
\beq 
\left( \begin{array}{c} \Phi(t,t') \\ \Pi(t,t') \end{array} \right) = 
K(t,t') \cdot \left( \begin{array}{c} \Phi \\ \Pi \end{array} \right) + 
\g(t,t')
\eeq 
Instead of considering the three vector of operators $(\Phi,\Pi,\1)$, 
we have eliminated the trivial equation for the unity which leaves us
with an inhomogeneity. The elements of $K(t,t')$ and 
$\g(t,t')$ contain the remaining independent elements of 
$\cV^\dag(t,t') $. The coefficients of the matrix $K$ and the vector 
$\g$ are constraint by (\ref{s1}), to wit,
\beq 
\left( \begin{array}{c} \ph(t) \\ \pi(t) \end{array} \right) = 
K(t,t') \cdot \left( \begin{array}{c} \ph(t')  \\ \pi(t') \end{array} \right) + 
\g(t,t') \lab{comp1} 
\eeq
For briefness, we will use the symbolic matrix notations $ (A \cdot
B)(x,z) \equiv \int_y A(x,y) B(y,z), 
\, (A \cdot v)(x) \equiv \int_y A(x,y) v(y), \, (A^T)(x,y) \equiv
A(y,x) $ and $(v \otimes w)(x,y) \equiv v(x)w(y)$
whenever the meaning is unambiguous. 
If we shift the operators $\Phi,\Pi$ by their expectation values, the
constant $\g$ can be eliminated, and the product condition (\ref{s2})
gives rise to the relation
\beq 
Z(t) = K(t,t') \cdot Z(t') \cdot  K^T(t,t') , \lab{keq} 
\eeq 
where $Z$ contains the quantum widths, 
\beq 
Z := \left(  \begin{array}{cc} \ph_2 - \ph \otimes \ph & \frac{1}{2} w
- \ph \otimes \pi \\ 
 \frac{1}{2} w^T - \pi \otimes \ph & \pi_2 - \pi \otimes \pi
\end{array} \right) .
\eeq
In order to solve the condition (\ref{keq}), we remember that the
reduced time evolution operator $\cV(t,t')$ is a time ordered
product which by construction obeys a transitivity condition 
$\cV(t,t') = \cV(t,t'') \cV(t'',t')$. If we let $t''=0 $ it becomes
apparent that double time dependence factorizes into a product where 
the second factor is the inverse first operator at $t'$. Consequently,
we can also split the Heisenberg coefficients into
\beq
K(t,t') = A(t) \cdot A^{-1}(t').
\eeq
If we put back that form into the restriction (\ref{keq}), one finds
that the quantum widths factorize into
\beq 
Z(t) = A(t) \cdot A^T(t) . \lab{agl} 
\eeq
We emphasize that Eq.\ (\ref{agl}) does not put any restriction on the
 free choice
of the initial expectation values.  
$Z$ by construction being a symmetric
matrix, can always be diagonalized by a unitary transformation and
thus can always be written as $Z = R D R^T = R \sqrt{D} 
( R  \sqrt{D} )^T$.

Further conditions  can be obtained from compatibility of the time
evolution (\ref{comp1}) with the equations of motion. For the very
general class of Hamiltonians composed of a quadratic kinetic part 
$\Pi^2/2 $ and a canonical momentum--independent potential $V(\Phi)$, the
equations of motion (\ref{eom1}) for
the position coordinate are exactly $\dot \ph(t) = \pi(t)$ and do not
get any further dissipative corrections. It follows in this case that 
$A$ has to have the structure  
\beq 
A(t) = \left( \begin{array}{cc} \a(t) & \b(t) \\ 
\dot \a(t) & \dot \b(t)   \end{array} \right) \lab{adgl} .
\eeq 
Together with (\ref{agl}) that form also implies 
$2 \dot \ph_2(t) = w(t) + w^T(t) $
which is in fact also exactly valid for theories with quadratic kinetic
terms. For Hamiltonians with a different kinetic term, one has to keep a
more general form of the matrix $A(t)$ which renders the algebra more
complicated, but the construction still works along the same
lines.  

A solution of the Eqs.\ (\ref{agl}-\ref{adgl}) can be
parameterized by $\a(t) + i \b(t) =$$ X(t) \cdot U(t)$, where
$U(t)$ has to be a unitary matrix and $X(t)$ be real. When plugging in 
that, we benefit from the fact that unitary matrices can be regarded 
as solutions of the first order differential equation $\dot U(t) = i
\Om(t) \cdot U(t)$ with $\Om(t)=\Om^T(t)$ being real. We get 
\beq 
\ph_2 - \ph \otimes \ph = X \cdot X^T , \quad 
\frac{1}{2} w - \ph \otimes \pi = X \cdot P^T \lab{XP}
\eeq 
with $P = \dot X$,  and 
\beq 
\pi_2 - \pi \otimes \pi - P \cdot P^T 
= X \cdot \Om \cdot \Om \cdot X^T . \lab{Om}
\eeq 
These relations can be combined into the matrix evolution operator,
which first row elements are found to read
\bea  \a(t,t') & = &  X(t) \cdot \cC(t,t')  \cdot X^{-1}(t') - \nn \\ 
&&  \b(t,t')  \cdot P(t')  \cdot X^{-1}(t') 
\eea 
and 
\beq \b(t,t') = X(t)  \cdot \cS(t,t')  \cdot \Om^{-1}(t') 
\cdot X^{-1}(t').
\eeq

The functional real generalisation $\cC,\cS$ of the trigonometric 
functions emerge from biquadratic expressions of the real and
imaginary part of the matrix $U(t)$ and solve the differential
equations 
\beq 
\dot \cC(t,t') = - \Om(t) \cdot \cS(t,t') , \quad 
\dot \cS(t,t') =  \Om(t) \cdot \cC(t,t') \lab{SC}
\eeq 
with initial conditions $ \cC(t,t) = 1  , \quad \cS(t,t) =0 $.
The complete form of the  matrix elements of the reduced evolution
operator can now be expressed in terms of the quadratic observables
and one action variable $\Om$.

\section{Equations of motion}

To be specific, let us study a model with Hamiltonian operator
$H=H_0+H_I$ with dynamically free part
\beq 
H_0 = \frac{1}{2} \int\limits_z ( \Pi(z)^2 +  \Phi(z)  \di_z  \Phi(z) )
\eeq 
containing the positive definite local operator $\di_z = m^2 -
\triangle_z$,  and interaction
\beq 
H_I = \frac{\la}{2} \int_z \Phi^4(z)  .
\eeq 
For that particular Hamiltonian, the equations of motion (\ref{eom1}) 
evaluate to the macroscopic evolution equations
\bea
\dot \ph(x) &=& \pi(x) ,  \lab{m1}\\
\dot \pi(x) &=& - \di_x \ph(x) + 2 \la \ph(x)  \left( 2 \ph^2(x) -  3  
\ph_2(x,x) \right) \nn \\   & &  + \la^2 \Si_\pi(x)  ,\\
\dot \ph_2(x,y) &=& \frac{1}{2} ( w(x,y) + w(y,x ) )  , \\ 
\dot \pi_2(x,y ) &=& - \frac{1}{2} \di_x w(x,y) + \la \left( 
4 \ph^3(x) \pi(y)- \right. \nn \\ && \left. 
3 \ph_2(x,x) w(x,y) \right) 
 + ( x \leftrightarrow y )  +  \nn \\
&& \la^2 \Si_{\pi_2}(x,y)  , \\ 
\dot w(x,y) &=& 2 \pi_2(x,y) - ( \di_x + \di_y ) \ph_2(x,y) \nn \\ 
 & & + 4 \la \left( 
2 \ph(x) \ph^3(y) - 3 \ph_2(x,y) \ph_2(y,y) \right) \nn \\ 
&& + \la^2 \Si_w(x,y) .\lab{m5}
\eea
where the expressions $\Si_\nu$ contain the dissipative
integral part. Analogously, the expectation value of the Hamiltonian 
is found to read
\bea  \bra H \ket_t &=& \nn \\ &&  \hspace{-2.7em}  
\frac{1}{2} \int\limits_z \left(  \pi_2(z,z) +\frac{1}{2} \di_z \ph_2(z,z)+ 
\la \left( 3 \ph_2^2(z,z) - 2 \ph^4(z) \right) \right. \nn \\ 
&& \quad \qquad \left. + \la^2 \Si_H(z) \right).
\eea 

The evaluation of the integral terms requires an approximation of the
projected evolution operator $T(t,t')$. 
By making the replacement,  $T(t,t') \to \cV(t,t') $ in (\ref{eom1}) 
we apply the first term of
a systematic approximation of the complete Liouvillian $L$
around the reduced evolution generated by $\hL(t)$.  
After acting with the conjugate of $\cV$ 
on the operators to its left which transforms them into the
generalized Heisenberg picture, we are left with averaged  operator
products which can be split into pairs by virtue of the generalized
Wick theorem. There, the propagator 
\bea
\lefteqn{ G(x,y|t,t')  =   2 ( X(t)  \cdot \cC(t,t') \cdot X^T(t')
)(x,y) }  \nn \\ 
& & \quad  =   \bra [  \Phi(x|t,t') - \ph(x|t) ] [ \Phi(y) - \ph(y|t') ] +
\mbox{h.c.} \ket_{t'}  
\eea
appears in a natural way.  

The dissipative expressions have the form of memory integrals,
\beq 
\Si_\nu(t) = \int\limits_z \int_0^t dt' \s_\nu(t,t') 
\eeq
where, after some lengthy algebra, the integrands are found to read 
\newcommand{\lef}{\hspace{-3.4em} }
\newcommand{\squad}{\quad \hspace{-6.3pt} }
{ 
\bea
\s_\pi(x) &=& 6  \b(x,z) \ph'(z) \left( 3 G^2(x,z) - \b^2(x,z) \right), \\
\s_{\pi_2} (x,y) &=& \nn \\ 
 && \lef  3  \left[  
\dot \b(y,z) G(x,z) \left( G^2(x,z) - 3\b^2(x,z) \right) + \right. 
\nn \\ && \lef \squad 
\b(x,z)  \dot G(y,z ) \left( 3 G^2(x,z) -\b^2(x,z) \right)  +
 {} \nn
\\ && \lef  \squad 6 \left( \dot \b(y,z) ( G^2(x,z) - \b^2(x,z) ) \right. +
\nn \\ && \lef \squad
\left. \left. 
{} 2 {} \b(x,z)  G(x,z) \dot G(y,z) \right) \ph(x) \ph'(z) \right] + 
\nn \\ && \lef \s_\pi(x) \ph(y) +  ( x \leftrightarrow y ) ,\\ 
\s_w(x,y) &=& \nn \\ 
&& \lef 6 \left[ 
\b(y,z) G(x,z) \left(3 G^2(y,z) - \b^2(y,z) \right) + \right. 
\nn \\ && \lef \squad 
\b(x,z) G(y,z) \left( G^2(y,z) - 3 \b^2(y,z) \right) + 
\nn \\ &&  \lef \squad 
2 \b(y,z) \left( 3 G^2(y,z) - \b^2(y,z) \right) \ph(x) \ph'(z) +
\nn \\ && \lef  \squad 
6 \left(  \b(x,z) G^2(y,z) + 2 \b(y,z) G(x,z) G(y,z)  \right. 
\nn \\ && \lef \qquad\qquad\qquad\qquad\qquad  \qquad \qquad \left.
- \b(x,z) \b^2(y,z) \right)  + 
\nn \\ 
&& \lef \squad  \left. 6 \left( \b(y,z) G(x,z) - \b(x,z) G(y,z) \right) 
\ph_2(y,y) \ph_2'(z,z)  \right], \nn \\ \\ 
\s_H(x) &=& \nn \\ 
&& \lef  
 - 3  \b(x,z) \left( G(x,z) ( 
 G^2(x,z) - \b^2(x,z) ) + \right. \nn \\ 
&& \lef  \left. 2 \ph(x) \ph'(z) ( 3  G^2(z,x) - \b^2(x,z) ) \right) 
\eea
}
with $\ph'(x) \equiv \ph(x|t'), \, \ph_2'(x,y) \equiv \ph( x,y|t')$ and
other dependencies on time arguments have been suppressed.       
The Eqs. (\ref{m1}-\ref{m5}) together with these relations finally 
form the closed set of macroscopic equations of motion. 

\section{Effective Hamiltonian} 

A natural question to ask is if the effective equations  of motion
which contain a memory term non--local in time still can be 
generated by a  macroscopic Hamiltonian or some action principle.
A straightforward candidate for that Hamiltonian 
is the expectation value $h=\bra H \ket_t$ which, however, has to be 
recast in terms of canonically conjugate variables.
 
Leaving aside the dissipative contributions for the moment,
the effective Hamiltonian was constructed systematically recently
\cite{nach98/1} for the zero--dimensional case. In analogy to that result, 
the equation of motion (\ref{m1}) suggest to introduce the
macroscopically canonically conjugate pair $\lbrace \ph(t) , \pi(t)
\rbrace  $. 
Furthermore, if we chose the positional uncertainty $X(x,y|t)$ as 
bilocal second position coordinate, the corresponding canonically
conjugate pair is found to be $\lbrace X(x,y|t),P(x,y|t) \rbrace $. 
Together with the relations
(\ref{XP},\ref{Om}), we find that the non--dissipative part of the
energy is composed of three contributions, $h_1 = T + V_{pot.}+
V_{am.} $, with a kinetic term 
\beq 
T = \frac{1}{2} \int\limits_z \left( \pi(z) \pi(z)  + 
( P \cdot P^T )(z,z) \right) , 
\eeq 
a potential
\bea V_{pot.} & = &
\frac{1}{2} \int\limits_z \left( \ph(z) \di_z \ph(z) + \frac{1}{2}
\di_z ( X \cdot X^T )(z,z) + \right.   \nn \\  && \quad   
 \left. 
\la \left(  3  ( X \cdot X^T)(z,z) ^2 + 
6 \ph^2(z) (X \cdot X^T)(z,z)   + \right. \right. \nn \\ 
&& \quad \quad \quad  \left. \left.  \ph^4(z)  \right) \right) ,
\eea
and an angular momentum term ($\cA \equiv X^T \cdot X \cdot \Om $), \\ 
\beq 
V_{am.} =  \frac{1}{2} \int\limits_z \left( 
( {X^T}^{-1}\cdot \cA \cdot \cA^T 
\cdot X^{-1} ) (z,z)
\right) .
\eeq
That term comes from the kinetic part of $\bra  H\ket_t$
when transforming into canonical coordinates \cite{nach98/1,coop97}
 
The Hamiltonian equations of motion for the variables $\ph,\pi ; X,P$ 
as being generated by $h_1$ 
are  equivalent to the non-dissipative parts of the original 
Eqs. (\ref{m1},\ref{m5}). This system, however, has the 
additional integral of motion $\cA$. 
In order to reproduce that property by the
effective Hamiltonian too, 
we introduce a canonically conjugate variable for the
momentum $\cA$. The equations of motion for the conjugate pair
$\lbrace \Psi(t) ,\cA(t) \rbrace $ 
then in fact yields a constant uncertainty $\dot \cA(t) =0$
since the non--dissipative Hamiltonian is independent of $\Psi$.

This canonical structure can be carried over to the dissipative
corrections. To see that in detail, we first realize that the energy
corrections $\Si_H$ do not depend on the momentum $\cA(t)$, but only on 
the quantities $\ph,\pi,X,P,\cS$ and $\cC$. Regarding the expectation
value for the energy   
as functional of those canonical variables, it follows
that 
\beq 
\dot \Psi = \frac{\d h }{\d \cA} = \Om  .
\eeq
By comparison with the definition  (\ref{SC}), it is possible to
split off the factor $\dot \Psi$ if we regard the generalized 
trigonometric functions as functionals of $\Psi$ being subject to the
defining equations 
\bea  \frac{\de \cC[\Psi] (u,v)}{\de \Psi(x,y) } & = &  - \de(x,u)
\cS(y,v) ,
\nn \\ 
\frac{\de \cS[\Psi] (u,v)}{\de \Psi(x,y) } & = &  - \de(x,u) \cC(y,v) .
\eea 
with initial conditions $\cC[\Psi =0 ] = \d , \, \cS[\Psi=0] = 0 $. 
The complete macroscopic Hamiltonian equations of motion with canonical pairs
$\lbrace \ph,\pi;X,P; \Psi, \cA \rbrace $ are thus generated by the classical
effective macroscopic energy functional
\beq 
h[\ph,\pi;X,P;\Psi,\cA]  = T + V_{am.}+ V_{pot.} + \la^2 \Si_H 
\eeq
and are in fact equivalent to the original equations of motion
(\ref{m1}-\ref{m5}).  
The effective macroscopic Hamilton functional is non--local in time,
and accounts for the full history of the system. The corresponding
dynamical initial value problem is only defined for the initial time
$t=0$ where the initial preparation is determined by Janynes'
principle of maximum uncertainty. The Hamiltonian $h$, however, cannot be
generated as classical limit of a corresponding microdynamical
Hamiltonian, since the reduced microdynamics is a Liouvillian
one. We further remark that the relevant entropy defined in 
(\ref{srel}) is automatically
larger or equal at later times than its value at $t=0$, though 
intermediate entropy fluctuations are not excluded from dynamics. 
We mention that by means of a standard Legendre
transform one straightforwardly gets the corresponding macroscopic 
Lagrangian and effective action. 

\section{Conclusion} 

We constructed the effective macroscopic equations of motion and
Hamiltonian for the statistical  ensemble defined by expectation
values and equal time correlators of position and momentum operators
in a scalar theory with quartic interaction potential. The systematic
construction based on the exact reduced equations of motion for the
corresponding level of observation have to be approximated
self--consistently in order to avoid violation of energy
conservation and for compatibility with the Robertson--equation of
motion. We solved that problem by a systematic approximation method based
on an effective microdynamic Liouvillian evolution compatible with time
evolution of the accompanying statistical operator, rather than using
an effective microdynamical Hamiltonian which fails to satisfy the
consistency requirements. The method fully accounts for entropy
generation and thus can be regarded as suitable tool to study the
effects of  dissipative time evolution even at arbitrarily 
large time scales. Moreover, the method intrinsically is superior to
the study of the BBGKY--hierarchy of equations of motion since it both
includes a systematic principle to solve the initial preparation
problem, and avoids the unresolved difficulty of how to truncate the
BBGKY--hierarchy without implicitly introducing approximation method
artifacts. Further applicaltions are planned for future research. 

\acknowledgments
I would like to thank U. Heinz and the local organizers of
the ``5th International Workshop on Thermal Field Theories and
Their Applications'' for all their efforts.

\end{document}